# The DAME/VO-Neural infrastructure: an integrated data mining system support for the science community

M. Brescia[1,2], A. Corazza[3], S. Cavuoti[2], G. d'Angelo[2], R. D'Abrusco[2], C. Donalek[4], S.G. Djorgovski[4], N. Deniskina[2], M. Fiore[3], M. Garofalo[2], O. Laurino[2], G. Longo[1,2], A. Mahabal[4], F. Manna[3], A. Nocella[2], B. Skordovski[2]

[1]INAF-OACN, National Institute of Astrophysics, Naples, Italy
[2]Department of Physic Science, University Federico II, Naples, Italy
[3]Department of Computer Science, University Federico II, Naples, Italy
[4]CALTECH, Californian Institute of Technology, Pasadena, California USA

*Abstract* — Astronomical data are gathered through a very large number of heterogeneous techniques and stored in very diversified and often incompatible data repositories. Moreover in the e-science environment, it is needed to integrate services across distributed, heterogeneous, dynamic "virtual organizations" formed by different resources within a single enterprise and/or external resource sharing and service provider relationships. The DAME/VO-Neural project, run jointly by the University Federico II, INAF (National Institute of Astrophysics) Astronomical Observatories of Napoli and Trieste and the California Institute of Technology, aims at creating a single, sustainable, distributed e-infrastructure for data mining and exploration in massive data sets, to be offered to the astronomical (but not only) community as a web application. The framework makes use of distributed computing environments (e.g. S.Co.P.E.) and matches the international IVOA standards and requirements. The integration process is technically challenging due to the need of achieving a specific quality of service when running on top of different native platforms. In these terms, the result of the DAME/VO-Neural project effort will be a service-oriented architecture, obtained by using appropriate standards and incorporating Grid paradigms and restful Web services frameworks where needed, that will have as main target the integration of interdisciplinary distributed systems within and across organizational domains.

## I. INTRODUCTION

Modern scientific data (such as in the Astrophysics environment) mainly consist of huge datasets that are gathered by a very large number of techniques and stored in very diversified and often incompatible data repositories. More in general, in the e-science environment, it is considered as a critical and urgent requirement to integrate services across distributed, heterogeneous, dynamic "virtual organizations" formed by different resources within a single enterprise and/or external resource sharing and service provider relationships. One first and important step in this direction has already been undertaken within the Astrophysics community with a set of initiatives that have flourished under the generic name of Virtual Observatory (VObs). The VObs, organized worldwide by means of the International Virtual Observatory Alliance (IVOA), has defined a set of standards to allow interoperability among different archives and databases in the astrophysical domain, and keeps them updated through the activity of dedicated working groups. One of the first main goals of the IVOA is the federation under common standards of all astronomical archives available worldwide. The concept is that having this meta-archive completed, its exploitation allows a new type of multi-wavelength, multi-epoch science which can only be barely imagined, but also poses unprecedented computing problems. So far, up to now most of the implementation effort for the VObs has concerned the storage, standardization and interoperability of the data together with the computational infrastructures. In particular it has focused on the realization of the low-level tools and on the definition of standards. Our project extends this fundamental target, by integrating it in a service oriented infrastructure, including the implementation of advanced tools for Massive Data Sets (MDS) exploration, soft computing, data mining (DM) and Knowledge Discovery in Databases (KDD).

*A. The theoretical domain*

From the scientific point of view, the DAME/VO-Neural project arises from the astrophysical domain, where the understanding of the universe beyond the Solar System is based on just a few information carriers: photons in several wavelengths, cosmic rays, neutrinos and gravitational waves. Each of these carriers has it peculiarities and weaknesses from the scientific point of view: they sample different energy ranges, endure different kinds and levels of interference during their cosmic journey (e.g. photons are absorbed while charged Cosmic Rays (CRs) are deflected by magnetic fields), sample different physical phenomena (e.g. thermal, non thermal and stimulated emission mechanisms), and require very different technologies for their detection. So far, the international community needs modern infrastructures for the exploitation of the ever increasing amount of data (of the order of PetaByte/year) produced by the new generation of telescopes and space borne instruments, as well as by numerical simulations of exploding complexity. More in detail, basic astrophysical requirements can be summarized in two items:

- The need of a "federation" of observed and simulated data, by collecting them through several worldwide archives and by defining a series of standards for their formats and access protocols;
- The implementation of innovative computing tools for data exploration, mining and knowledge extraction, user-friendly, scalable and as much as possible automatic;

In other words, the incoming availability of multi-band and multivariate data, from ever more accurate observations of the Universe, introduces the urgency to adopt more powerful soft computing methods to explore from different perspective the MDS. Having such DM tools available, it could be possible to go deeply into the investigation of main astrophysical problems, like:

- Detection and study of the photometric (variable objects) and astrometric (Near Earth Objects or NEO) transients in archival and digital survey data;
- Physical classification of the extra-galactic objects paying special care to the spectroscopic classification of Active Galactic Nuclei (AGN);
- Better knowledge on AGN physical properties;
- Star/Galaxy separation and classification;
- Automatic evaluation of the Point Spread Function (PSF) in frames coming from digital surveys;
- Auto-adaptive integration of spectroscopic and photometric data, such as the evaluation of photometric redshifts as generalization of feature learning on spectroscopic data;
- Integrated modeling of extremely large telescopes and complex focal plane instruments, through simulation pipelines, covering optical, Finite Element Analysis (FEA), mechanical and control design aspects;

All these topics require powerful, computationally distributed and adaptive tools able to explore, extract and correlate knowledge from multivariate MDS in a multi-dimensional parameter space. The latter results as a typical data mining requirement, dealing with many scientific, social and technological environments.

Concerning the specific astrophysical aspects, the problem, in fact, can be analytically expressed as follows:

Any observed (or simulated) datum defines a point (region) in a subset of $R^N$, such as:

- R.A. and DEC;
- time and $\lambda$;
- experimental setup (i.e. spatial and/or spectral resolution, limiting magnitude, brightness, etc.);
- fluxes;
- polarization;
- spectral response of the instrument;
- PSF;

Every time a new technology enlarges the parameter space or allows a better sampling of it, new discoveries are bound to take place.

So far, the scientific exploitation of a multi-band (D bands), multi-epoch (K epochs) universe implies to search for patterns and trends among N points in a DxK dimensional parameter space, where $N > 10^9$, $D >> 100$, $K > 10$.

The problem also requires a multi-disciplinary approach, covering aspects belonging to Astronomy, Physics, Biology, Information Technology, Artificial Intelligence, Engineering and Statistics environments. In this sense, experimental astronomy has in practice become a three players game, made by:

- Astronomers: theory, data, understanding, discoveries, structure, biases;
- Statisticians: evaluation of data, validation, analysis, dimensional reduction, models;
- Computer Scientists: implementation of infrastructures, archives and databases, information retrieval, middleware, scalable tools, OOP;

*B. Data Exploration*

The requirements coming from above theoretical scenario involves a data exploration conceptual domain based on the approach summarized in Fig. 1.

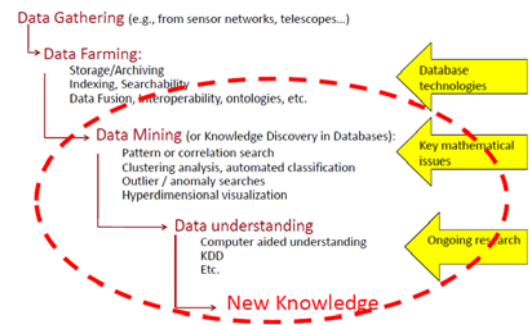

Fig. 1 – data exploration conceptual domain

In the last decades data exploration research has been driven by the advance of database technology and the creation of huge datasets. The growth of "visual and/or numerical analytics", has emphasized the danger of information overload and the need to harness new technology, such as high resolution and multivariate analysis systems, in order to maintain overview and control of the data. DAME/VO-Neural group have always stressed the need to organize in an homogeneous way both syntax and semantics of data manipulation, analysis and reduction, with the aim at the creation ontologies to ease communication between scientists using a distributed computing infrastructure such as the S.Co.P.E. GRID, representing environment architecture which is main deployment target of DAME/VO-Neural Project. Despite these advances, the problem of finding suitable methods for the exploration of large datasets remains still formidable. The emphasis is on methods of bridging the gap between accurate representations and mining of the data and the capabilities of current technology and users. The analytical methods based partially on statistical random choices (crossover/mutation) and on knowledge experience acquired (supervised and/or unsupervised adaptive learning) could realistically achieve the discovery of hidden laws behind focused phenomena, often based on nature laws, therefore the simplest.

During the R&D phase of our project, aimed at define and characterize rules, targets, ontologies, semantics and syntax standards, the following functional breakdown structure has been derived:

- Exploration
- Dimensional Reduction
- Regression
- Clustering
- Classification
- Visualization

It provides a taxonomy between possible data exploration modes, made available by our infrastructure as data mining experiment typology (use case). In order to accelerate the infrastructure prototype development, in the first implementation phase it was decided to reduce this scenario, by focusing the attention on Classification and Regression functionalities only.

*1) Data Exploration: Classification*

Statistical classification is a procedure in which individual items are placed into groups based on quantitative information on one or more features inherent to the items (referred to as features) and based on a training set of previously labelled items. A classifier is a system that performs a mapping from a feature space X to a set of labels Y. Basically a classifier assigns a pre-defined class label to a sample. Formally, the problem can be stated as follows: given training data $\{(x\_1,y\_1),...,(x\_n, y\_n)\}$, (where x_i are vectors), a classifier h:X->Y maps an object x ε X to its classification label y ε Y.

Different classification problems could arise:

a) <u>crispy classification</u>: given an input pattern x (vector) the classifier returns its computed label y (scalar).

b) <u>probabilistic classification</u>: given an input pattern x (vector) the classifier returns a vector y which contains the probability of y_i to be the "right" label for x. In other words in this case we seek, for each input vector, the probability of its membership to the class y_i (for each y_i).

Both cases may be applied to both "two-class" and "multi-class" classification.

*2) Data Exploration: Regression*

We will define Regression as the supervised search for a mapping from a domain in $R^n$ to a domain in $R^m$. One can distinguish between two different types of regression:

a) <u>data table statistical correlation</u>: the user tries to find a mapping without any prior assumption on the functional form of the data distribution. Machine learning algorithms are well suited for this kind of regression;

b) <u>function fitting</u>: with curve fitting the user tries to validate the hypothesis, suggested by some theoretical framework, that the data distribution follows a well defined, and known, function;

A regression system performs a mapping from a parameter space X to a target space Y. Formally, the problem can be stated as follows: given training data $\{(x\_1,y\_1),...,(x\_n, y\_n)\}$ (where x_i are vectors) a regression operator h:X->Y maps an object x ε X to its value y ε Y.

## C. The Project Infrastructure

By taking into account both theoretical and conceptual domains, we designed an infrastructure based on the following skill features:

- ✓ Object Oriented Programming;
- ✓ Internal standards and protocols (VO, XML);
- ✓ Java language (almost generic for data mining models);
- ✓ User/Session Registry DataBase Management System (MySQL);
- ✓ Web-based User I/O;
- ✓ Service-Oriented Web Application and Restful Web Service Technology, Servlet (Web Server applets);
- ✓ Plugin-based Modularity (easy to be integrated/modified) for data mining models;
- ✓ Hardware independent through GRID driver;
- ✓ Data conversion and manipulation support (ASCII, FITS, CSV, VOTable);

In the following sections, each of the infrastructure components is described.

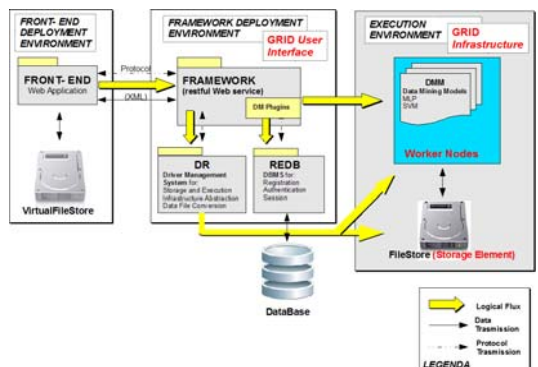

Fig. 2 – DAME/VO-Neural Infrastructure Architecture

### 1) Front End Component

The Front End (FE) represents the main interface between the infrastructure core and the external world (user who wants to submit own experiments). It is structured as a Service-Oriented (SO) tool implemented by a WEB application, whose scope is to furnish a I/O data and functionalities computing interface to the user. This choice has been driven taking into account following requirements:

- To provide a remote distributed application, whose internal mechanisms are totally hidden to the final user;
- To avoid any HW/SW client dependence. It is possible to use any kind of public web browser to interact with the WEB application in a dynamical (context-dependent) way;
- To implement an easy and safe security policy, handled by the Servlet (Web Server applet);
- To obtain an easy bug-free and maintainable application;
- To provide a flexible instrument easy to be updated and expanded;

A registered and authenticated user can access the FE via browser in order to fill in the login form obtained from the previous registration. In case of deployment of the infrastructure on a GRID platform, a robot certificate system is automatically provided in a transparent way to the final user. The GUI of the FE permits the navigation of the user between his own working sessions (term introduced to classify user experiments, i.e. associations between chosen datasets, data mining model and functionality), the remote upload of user data files, their interactive manipulation (preparation of data sets for the specific data mining experiment) and storing for future experiments. The FE also provides the interactive/asynchronous check of experiment status, the download of final or partial results and the data mining model-dependent output graphics and charts eventually selected by the user.

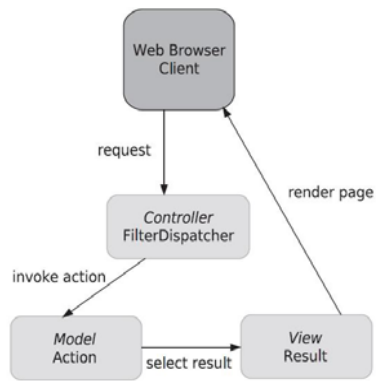

Fig. 3 – The FE MVC pattern implementation

Any user request is internally processed through the XML-based bi-directional communication between FE and the core component (FrameWork).

The FE internal mechanism is based on the Model-View-Controller (MVC) pattern that provides a separation of concerns that applies well to WEB applications. Separation of concerns allows us to manage the complexity of large software systems by dividing them into high-level components.

The MVC design pattern identifies three distinct concerns: *model*, *view*, and *controller*. In the FE, these are implemented by the Action, Result, and Filter Dispatcher respectively (Fig. 3).

*2) FrameWork Component*

The DAME/VO-Neural Suite is composed of several components which relies on a common infrastructure to perform their task. This infrastructure is basically provided by the FrameWork (FW) component. Nonetheless this component, while providing this common platform, represents the core of the project, since it is the component who actually implements, configures and launches the data mining experiments. In our analysis of the FW component, every operation can be considered atomic, so the natural choice is to implement the web service in the RESTful architectural style. A RESTful Web Service is a client\server architecture in which the web services are viewed as resources and are identified by their URLs. A client that wants to access to a resource can use common http methods like get and post on the URL that identifies the desired resource. A RESTful environment is completely stateless, that is every interaction is atomic. The main advantage of this style is its great simplicity, which allows a fast deployment and simple interactions with the external world.

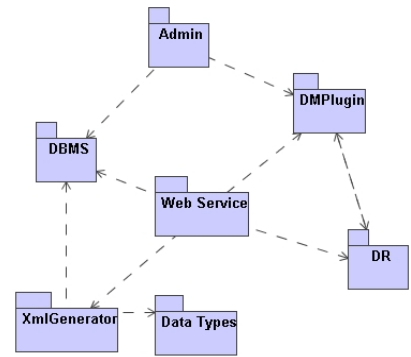

Fig. 4 – The FW class diagram

We identified Servlets (namely an object which accept a request from a client and creates a response based on that request) to implement the RESTful Web Service. For every resource there will be a Servlet associated to it, which overrides the standard methods to implement the operation associated to that resource. Furthermore, the FW component needs to implement an Admin interface. An Admin needs to interact with the FW to add or remove plug-ins (DMMPlugin) and possibly to request statistical information like number of users and number of experiments. Moreover, one of the most critical factors of the FW is the interaction of a newly created experiment with the GRID platform. The FW component needs to create and configure the DMPlugin associated to the experiment. We called DMPlugin a module encapsulating experiment and associated Data Mining (DM) configuration and working flow methods. After such a plug-in is configured it needs to run the experiment by calling the run method of the plug-in itself. In GRID, as known, the process needs to migrate on a Worker Node (WN). To implement this migration we've chosen to serialize the DMPlugin in a file. Serialization is a process of converting an object into a sequence of bits so that it can be stored on a storage medium. To manage the interactions between the FE and the FW the adopted standard protocol, in some operations like the creation of a new experiment, is based on XML-type files. The FW and the DMPlugin have methods to create and parse XML files to apply user (FE) requests.

### 3) REgistry & DB Component

The Registry & DataBase (REDB) is the Suite component including the DataBase Management System (DBMS) component and its Interface. The REDB handles user's registration, authentication, working sessions, experiments and files; its DBMS manages data and relationships about Users, Sessions, Experiments, Functionalities and Files. The component architecture is based on the following entities: a relational DBMS composed by a DB Server and a text interface Client that supports the SQL syntax, a DBC Driver, a DBC API and the REDB Interface.

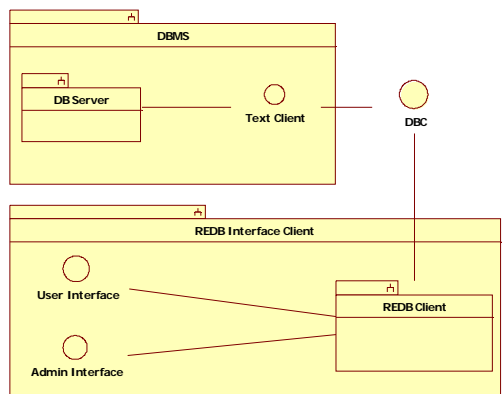

Fig. 5 – The REDB architecture

The component architecture is based on the following technologies: MySQL, Java as interface programming language, JDBC Driver and its API (Fig. 6).

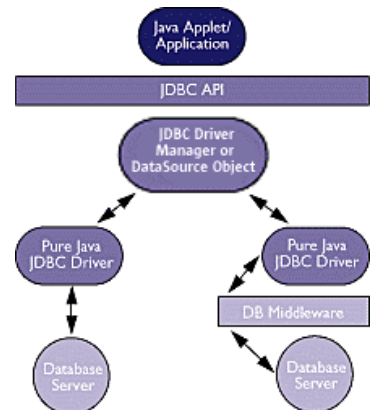

Fig. 6 – The JDBC driver architecture

### 4) DRiver Component

To separate FW functional requirements from their implementation issues, it was necessary to create a library of methods called DRiver (DR) Management System. The DR is the component used by the FW to manage the processing environment. It implements the low-level interface with computational environment, in order to permit the FW implementation, through the specific drivers, on different platforms (such as Stand-Alone or GRID). In other words, the DR is used to implement the proper access to the platform-dependent resources required by all the specific use cases and functionalities of the Suite. The experiment I/O data basically consist in the user input and suite output. In order to avoid multiple deployment of these data on several components, it was decided to provide a file storage system (SE or Storage Element on GRID), hosting real data files, handled directly by the FW through the DR component methods. This component includes also a library of data file format (FITS, ASCII, CSV and VOTable) translating methods, used by the FW depending on the specific DM model data format supported.

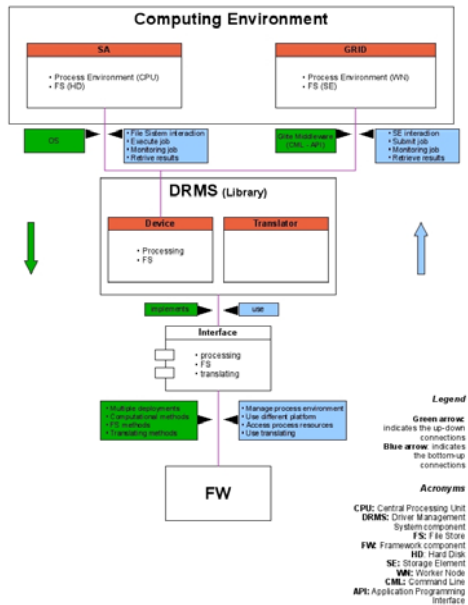

Fig. 6 – The DR component architecture

*5) Data Mining Models Component*

The DMM is the component that implements the DM models and related wrappers code and their use cases.

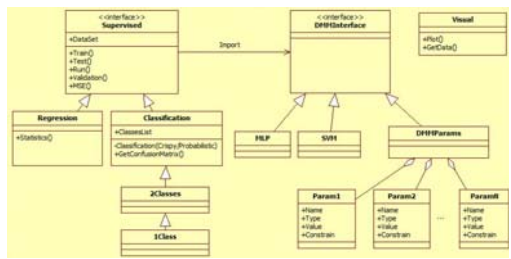

Fig. 7 – The DMM component architecture

Its main features can be summarized as follows:

1. Implementation oriented to the functionalities (i.e. classes of functionalities, such as Classification and Regression);
2. Possibility of use functionalities with more than one model without duplicate code (Pattern Bridge as standard design pattern);
3. Separation between Supervised and Unsupervised models;
4. A common interface for all the models (by means of a specific class rendering typical data mining self-adaptive parameters, called DMMParams);

In the first release, the DMM implements MultiLayer Perceptron (MLP) ([9]) and Support Vector Machine (SVM) ([10]), as supervised data mining models, while Self Organizing Maps (SOM) ([9]) is the unsupervised model foreseen to be implemented in the next Suite release.

II. CONCLUSION

The DAME/VO-Neural project comes out as an astrophysical data exploration tool, originating from the very simple consideration that, with data obtained by the new generation of instruments, we have reached the physical limit of observations (single photon counting) at almost all wavelengths. Hence, our opportunity to gain new insights on the Universe will depend mainly on the capability to recognize patterns or trends in the parameter space, which are not limited to the 3-D human visualization, from very large multi-wavelength and multi-technique data sets. Moreover, this approach (and of course the proposed technique) can be easily and widely applied to other scientific, social, industrial and technological scenarios. Our project has recently passed the R&D phase ([1], [2]), de facto entering in the implementation step and by performing in parallel the scientific testing with first infrastructure prototype, accessible, after a simple authentication procedure, through the official project website address (http://voneural.na.infn.it/). First scientific test results ([5], [6], [8]), confirm the goodness of the theoretical choice, of the data mining model design and technological strategy. This demonstrates and encourages the winning approach to integrate VObs experience, massive data exploration, Knowledge Database

Discovery, GRID paradigm, Soft Computing, Service-Oriented Information Technology in a unique infrastructure for the entire community.

ACKNOWLEDGMENTS

The authors would like to thank VOTECH and S.Co.P.E. projects for their fundamental support, together with all DAME/VO-Neural team members, contributors and collaborators for their effort employed to reach the successful goal. The project was funded by the Italian Ministry of Foreign Affairs (MAE) through a bilateral Italy-USA agreement.

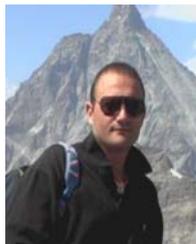 **M. Brescia** Astronomer researcher at INAF-OACN (Astronomical Observatory of Capodimonte of Naples) and member of (IAU) International Astronomical Union. Degree in Computer Science, specialized in astrophysical technology, artificial intelligence and data mining software engineering. Teacher of Computer Architecture at Computer Science Department and Astronomical Technologies at the Physics Department of the University Federico II, Naples, Italy.

A**. Corazza** assistant professor at the University Federico II in Naples, Italy. Before, she worked at the ITC-irst (nowadays FBK) in Trento and at the University of Milan. Her research interests focus on statistical approaches to natural language processing, bioinformatics, and information retrieval